\documentclass[sn-basic]{sn-jnl}

\usepackage{graphicx}
\usepackage{multicol}

\usepackage{url}
\usepackage{booktabs}
\begin{document}

\title{Registered Reports in Software Engineering}

\author*[1]{\fnm{Neil A.} \sur{Ernst}}\email{nernst@uvic.ca}
\author[2]{\fnm{Maria Teresa} \sur{Baldassarre}}\email{mariateresa.baldassarre@uniba.it}

\affil[1]{\orgdiv{Department of Computer Science}, \orgname{University of Victoria}, \orgaddress{\city{Victoria}, \state{BC}, \country{Canada}}}

\affil[2]{\orgdiv{Dipartimento di Informatica}, \orgname{Università degli studi di Bari}, \orgaddress{\city{Bari},  \country{Italy}}}

\abstract{
Registered reports are scientific publications which begin the publication process by first having the detailed research protocol, including key research questions, reviewed and approved by peers. Subsequent analysis and results are published with minimal additional review, even if there was no clear support for the underlying hypothesis, as long as the approved protocol is followed.
Registered reports can prevent several questionable research practices and give early feedback on research designs. 
In software engineering research, registered reports were first introduced in the International Conference on Mining Software Repositories (MSR) in 2020. They are now established in three conferences and two pre-eminent journals, including this one (EMSE).
We explain the motivation for registered reports, outline the way they have been implemented in software engineering, and outline some ongoing challenges for addressing high quality software engineering research. 
}


\maketitle

\section{Introduction}
Registered reports are a model of scholarly publication which prioritize the importance of study design and significance rather than study outcomes. Focusing on whether the study was suitable to support the inferences of interest decouples publication from a focus on headline-worthy `significant' results.

In software engineering (SE) research, empirical methods are now standard.  
The top conferences in the field emphasize ``the extent to which the paper's contributions and/or innovations address its research questions and are supported by rigorous application of appropriate research methods.\footnote{\url{https://conf.researchr.org/track/icse-2022/icse-2022-papers?\#Call-for-Papers}}''

Sometimes these research methods are deployed as part of studies seeking to inductively (and occasionally abductively) \emph{explore} new insights into software engineering challenges. Other times empirical methods are used to deductively \emph{confirm} existing theories about the world. 
This is the distinction between exploratory vs. confirmatory research.
Other dichotomies in empirical SE research are also important, such as research strategy (simulation studies, interview studies, lab experiments, and others as outlined in ~\cite{Storey2020TheWW}), and perhaps most of all, the research's underlying philosophical perspective (the epistemic claims it believes it can make or should make). 

With empirical methods, however, can come undesirable side-effects that reduce confidence in the practical significance of the conclusions. These side-effects have been labelled as {\em questionable research practices} (QRPs, ~\cite{John2012}. 
QRPs occur when researchers are not clear about the type of study they conduct. 
For example, hypothesising after results are known (HARKing) is perfectly acceptable in an \emph{exploratory} context, but unacceptable in a confirmatory study, since deriving inferences having seen the results is an improper inference.
Similarly, the notion of forking paths, where researcher bias selects the most interesting results after seeing the data~\citep{Simmons2011FalsePositivePsychology} is a useful way of highlighting potentially significant results for future studies. Forking paths characterize exploratory analyses of different machine learning configurations, for example. 
A more detailed description of these issues in software engineering can be found in \cite{Neto2019EvolutionOS}.

Common to these problems is an insistence on the significance \textit{of the results} as publication criteria, rather than the importance of the question and soundness of the method. For example, if we compare code quality as a dependent variable in an experiment looking at the use of test-driven development (independent variable, TDD), should we not publish a result that finds insufficient evidence for a difference in the TDD vs not TDD treatment? Such a finding (from a well-conducted experiment) is still useful: it says there is no evidence TDD helps or hurts code quality. Practitioners would presumably be interested in this finding (at least as much as the counterpart, that TDD helps code quality)\footnote{see \cite{gharfari} for a more thorough discussion of TDD experiments}. Note too, that this is not the same as saying there is  no effect. 
Our aim should be well-conducted studies with a rigorous method, i.e., a study which \emph{could} find an effect if one was present. 


Registered reports (RR) help avoid this results-orientation because a RR approval shifts focus onto the soundness of the research plan and significance of the question. Publication ensues if the plan is followed, \emph{independent} of the actual results. Thus research efforts which fail to find an effect can be more common in publication \citep{Chambers_2020}. This happily reduces the bias in published results that impacts meta-analysis and systematic reviews (the file-drawer effect). 

\section{Background}
Post-hoc rationalizing is when researchers construct narratives to explain the data they found in a study. This story-telling \citep{Gelman2014} is an important aspect of science: it is the inductive/abductive aspect of theory building, and key to exploratory analysis where we seek to better understand the problem, or lay out plausible reasons why a solution worked. However, when researchers embark on theory-testing, or deductive, confirmatory research, they are using the collected data to test a theory. 

For example, we know enough about software development to believe that frequently changed (churned) files are more bug-prone. \cite{Nagappan2005} showed this was true at Microsoft. A confirmatory study might therefore look at testing this finding (what we might loosely call a theory) with new data (for example, in a startup company). In this confirmatory approach, looking at the data and reconstructing an explanation post-hoc is statistically invalid. 

Let us assume as researchers we adopt a Neyman-Pearson frequentist perspective (the vast majority of SE studies follow this perspective, at least implicitly). 
Let us further assume we followed other best practices in statistical inference, such as estimating the study's power to find an effect of a given size, and using a causal model with proper controls for colliders.
Then, we should only decide that the alternative hypothesis, i.e., that churn is predictive of defects, is either supported or not supported. 
Under a Neyman-Pearson approach `support' means that our long-run probability $p$ of observing the same data or more extreme values is less than some predefined $\alpha$, or, as Läkens writes, we are only likely to be misled if we assume the alternate hypothesis at most $\alpha \%$ of the time.\footnote{See Läkens's excellent course \url{https://lakens.github.io/statistical\_inferences/pvalue.html} for more on N-P inference.}

The issue with post-hoc rationalization in confirmatory research is that many explanations (forking paths or researcher degrees of freedom) can be found given a particular dataset~\citep{Gelman2014TheSC}. This means that estimating the true effect is impossible. Continuing the example, perhaps we look at the startup's data and decide that while churn did not predict defects, this is because the startup has a continuous delivery culture. Other explanations may be equally plausible though (perhaps we only looked at data from the best team). For such confirmatory studies, researchers should ensure that the study outlines its theory (including theoretical and practical estimands as outlined by \cite{Lundberg2021}) \emph{before} the data is collected and analyzed. Too often, such speculation---while entirely appropriate in science---is disguised as being supported by the statistical evidence from the study. 

Does software engineering research suffer from researcher bias problems like those mentioned above? Several studies report on the lack of statistical maturity in the literature \citep{Neto2019EvolutionOS}, for example, not tracking effect sizes \citep{Kitchenham2019MetaanalysisFF} or not referencing existing theories \citep{Hannay2007ASR}, for example, by creating causal models that outline constructs and context~\citep{Rohrer2018ThinkingClearlyAbout}.\footnote{Note these issues are often \emph{unconscious}, and not deliberate.}

One way to deal with researcher bias, already adopted in other fields, is pre-registration \citep{Chambers2013}. A pre-registered study is a research protocol, including planned hypotheses, data collection, and data analysis, that is announced---registered---before the study in full is conducted. This prevents post-hoc rationalizing (because the protocol has committed to the tests and expected outcomes) and the problems mentioned above. Registration is as simple as a blog post, or depositing an official document on a registry server, such as the ones supported by the Open Science Foundation\footnote{\url{https://cos.io/rr/}}.

\emph{Registered reports} expand on pre-registration by publishing the registration as a study plan. That plan is reviewed and approved by peers in a Stage 1, leading to ``in-principle acceptance” by a partner journal (such as EMSE). In principle acceptance means the journal commits to publish the study results \textit{even if} the results are not significant, assuming the study question is interesting, the study protocol sound, and the data collection adequate (examined in Stage 2). 
Evidence to date suggests RR helps improve study quality and scientific impact; for example, more RR studies are published which do not find effects~\citep{Chambers_2020}. 

\subsection{Related Efforts}
Work on open science efforts share some goals with registered reports, de-emphasizing novelty of the finding in favour of replications (or failed replications) of previous work, and studies that show no support for a well-founded hypothesis. Some examples include Replication and Negative Results (RENE) tracks at the International Conference on Software Analysis and Reverse Engineering (SANER), and an EMSE special issue on negative results \citep{Paige2017}.

The RoSE festival series\footnote{e.g., \url{https://github.com/researchart/rose-fse18}} initiated by Tim Menzies and others is about ``Recognizing and Rewarding Open Science in Software Engineering”. Open science principles and the idea of RR are in alignment: it is a key part of RR that protocols and results are shareable and public. Similar open science efforts such as the EMSE Open Science initiative\footnote{\url{https://github.com/emsejournal/openscience}} are likewise spiritual cousins of the RR efforts.
Registered reports were first introduced at the journal Cortex in 2013 although the idea of protocol review had been around earlier. \cite{Chambers_2020} provides a summary. Many journals now support the format, with the Empirical Software Engineering Journal (EMSE) supporting them as of 2020.
As of 2022, the ACM journal Transactions on Software Engineering and Methodology (TOSEM) hosts a Registered Paper initiative\footnote{\url{https://dl.acm.org/journal/tosem/registered-papers}}. It follows a journal-only model, not the conference+journal model described here.

\section{How It Works}
\begin{figure}
    \centering
    \includegraphics[width=.7\linewidth]{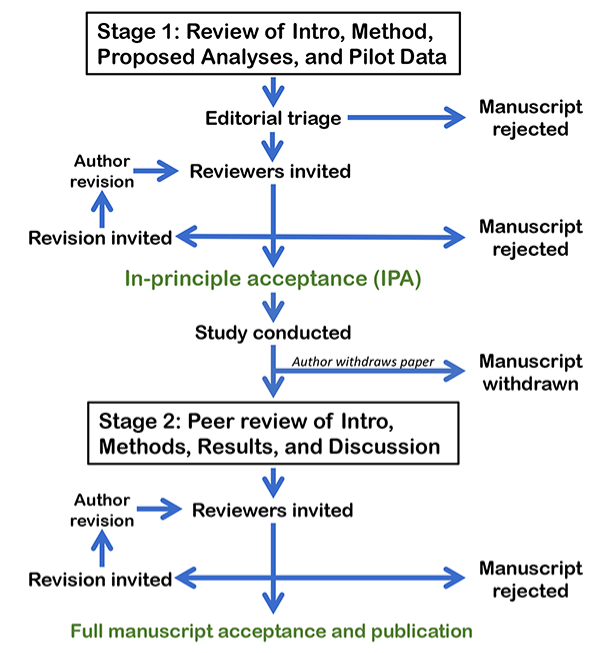}
    \caption{Stages of the Registered Reports workflow. Center for Open Science (\url{https://www.cos.io/initiatives/registered-reports?\#tabid3}) CC-BY-NoDerivs 4.0}
    \label{fig:rr-stages}
\end{figure}

Registered report studies follow a two-stage process with a workflow as in Fig. \ref{fig:rr-stages}.

\textbf{Stage 1: } Reviewers of the RR track review the submitted registered report. The modification from the typical RR approach at the Empirical Software Engineering Journal (EMSE) is that Stage 1 is managed as a conference track. Current options include the International Conference on Mining Software Repositories, the International Conference on Software Maintenance and Evolution, and the Empirical Software Engineering and Measurement Conference.

The submission for Stage 1 is usually 5-6 pages. It includes an introduction to the research topic and rationalization of the research questions/hypotheses, operationalization of variables, methodology and analysis pipelines. The research is evaluated for the novelty, importance, significance of the questions, and the soundness of the methods chosen (i.e., can they answer the question posed).
Where applicable, pilot data can also be submitted. 
Stage 1 is known as \textbf{in-principle acceptance (IPA)}. The Stage 1 report is typically posted to a preprint server such as ArXiv, although embargos are possible. 

\textbf{Stage 2:} Once a report has been accepted for Stage 1, the study is conducted and actual data collection and analysis takes place. In our community of software research, the report is also presented at the conference for comment. 
The results in Stage 2 can be negative! But if the protocol is adhered to (or minor deviations are thoroughly justified), the study is published. In practice, the Stage 2 review process has resulted in the first (journal) decision being a request for minor revisions, rather than (more typically) major revisions or even rejection.
Of course, this being a journal submission, a revision of the submitted manuscript may be necessary, as the participating journal (EMSE) maintains its quality standards. Reviewers will especially evaluate how precisely the protocol of the accepted pre-registered report is followed.

Complete review criteria based on the Open Science Foundation overview\footnote{osf.io/rr} is available as part of the SIGSOFT empirical standards initiative \citep{Ralph2021}\footnote{\url{https://github.com/acmsigsoft/EmpiricalStandards/blob/master/Supplements/RegisteredReports.md}}. Updates can be added via pull request.





\section{Early Lessons from RR}
Registered reports tracks have elicited Stage 1 submissions at MSR, ICSME, and ESEM, with more in the pipeline (see Table \ref{tab:stats}). To date (late 2022) six papers have successfully been published as completed Stage 2 reviews in EMSE, and 16 more are under Stage 2 review at EMSE.  

\begin{table}
    \begin{tabular}{ccccc}
        \hline
                   & \multicolumn{2}{c}{Stage 1}            & \multicolumn{2}{c}{Stage 2} \\
        Venue      & Submissions & IPAs & Submissions  & Publications \\ \hline
        MSR 2020   & 13          & 6                        & 4            & 3            \\
        MSR 2021   & 10          & 6                        & 4            & 1            \\
        MSR 2022   & 14          & 2                        & 1            & 0            \\
        ICSME 2020 & 7           & 4                        & 3            & 2            \\
        ICSME 2021 & n/a         & 6                        & 3            & 0            \\
        ESEM 2021  & n/a         & 4                        & 1            & 0            \\
        ESEM 2022  & 13          & 3                        & 0            & 0            \\ \hline
        \end{tabular}    
    \label{tab:stats}
    \caption{RR Submissions and Publications Since Inception at EMSE. Note that some studies were affected by the COVID-19 pandemic. Data may be incomplete as tracking submissions can be challenging.}
\end{table}

As part of our work on the registered reports track at MSR in 2020 (the first RR in a software conference), we ran a small survey with the participants and reviewers to assess the initiative. 
We received 25 responses. Most encouragingly, all participants would submit again to a RR track. Feedback addressed the report format, which followed an existing OSF guide and was not standard in SE research. Most participants (reviewers and authors) felt there was a lack of detail possible in four pages, or without a detailed pilot study. Finally, 18 respondents were comfortable with the Stage 1 acceptance leading to an EMSE paper in Stage 2, while 6 respondents were not comfortable with this.

Regarding the notion of In Principle Acceptance (IPA): ``[...] the fact that the results are missing, helps reviewers and authors focus on the methodological issue, which is a great added value in the review process [...]" and people appreciated that it helps reduce publication bias against negative results. But one reviewer noted: ``I felt a bit uncomfortable to have this burden on my shoulders as a reviewer so early in the process." Reviewers were aware that they were reviewing a paper that might get published in the top venue in SE, with expected high standards. Some reviewers and authors appreciated the way the Stage 1 reviews allowed for author rebuttals: ``I thought the entire goal was to help shape the methodology to be followed." But this back and forth is limited by the short cycles for conference publications, so there was some call for an extended discussion period. 
There was discussion in the survey responses, as well as among reviewers, about what was suitable as a registered report. We discuss this more in the next section. 

To improve, we had suggestions on page limits, writing guidelines, and the review tools: ``A more interactive pre-rebuttal stage so to speak". The respondents all agreed that the process was quite distinct from a full paper review, where the one of the key tasks of the reviewers is to advise editors if the paper is ready for publication. Instead, RR reviews focused much more on the scientific approach and protocol, which was heartening. But our existing tooling is designed for publication recommendation rather than interactive discussion. 

We have a few other insights from managing the overall process. The first is that the burden on editors/track chairs can be high. First, one must educate reviewers about the nature of RR tracks and the difference in criteria between Stage 1 vs Stage 2 reports (although this has been growing easier as the idea matures). Tool support for editorial duties can be challenging: it is hard to track all the studies as they bounce between multiple venues, sponsors, and tools (such as EasyChair, HotCRP, EditorialManager). This ``editorial tennis'' adds extra drag to the time to decision. If a reviewer drops out between Stage 1 and Stage 2, the new reviewer needs to begin from the start (or feels they do), slowing things. Page limits in Stage 1 might lead to a protocol missing important aspects that arise in Stage 2. 

Some of the deviations that occur to date include recruiting fewer participants than expected, or participants from slightly different pools. We have also seen deviations around study constructs. Constructs for measuring effects in SE practices can be difficult to define, such as the notion of productivity. In these cases, either reviewers accepted the justification, or the review process reverted to the full journal paper review common to EMSE non-RR submissions.

We have also devised a policy for conducting RR studies, to address questions about authorship and reviewer conflict (e.g., if a reviewer subsequently becomes conflicted in Stage 2 due to no longer being a blind review). Changes in authorship require a formal notification letter signed by all authors acknowledging the ACM/IEEE authorship criteria. Stage 1 reviewers and their students cannot become authors for ethical reasons. Stage 1 acceptance cannot be used to incentivize new project contributors. Finally, new conflict of interest checks are needed with new authors. New authors should be aware of how this can complicate reviewing.

A few other concerns expressed early on have not materialized: people submitting many Stage 1 proposals to get early feedback that a supervisor could have provided, or a Stage 1 submission being scooped by someone copying the protocol. This last merits some more discussion: our belief is that by the time Stage 1 is agreed and registered, it would be difficult to beat the authors to a Stage 2 result. We also support the notion of embargoed Stage 1 submissions, in the event this becomes a big concern, which registration tools on sites such as the Open Science Foundation also support. 

Pre-registration is in its infancy and subject to extensive philosophical debate. We refer the reader to the research dialogs in the Journal of Consumer Psychology for some point/counterpoint discussions about the value of registration (in particular, \cite{Pham2021PreregistrationIsNeither} and \cite{Simmons2021PreRegistrationIs}).

\section{Discussion}
\subsection{Three Benefits of Registered Reports}
Registered reports aim to provide early-stage feedback to authors and reduce researcher bias problems. In our experience with RRs at MSR, ICSME, ESEM and the journal EMSE, we think the following three items reflect different aspects of the RR process, and notably different benefits of using registration. We capture other benefits in Table \ref{tab:benefits}.
\begin{description}

\item[RRs offer early feedback on study design] The conference+journal form of RR used at EMSE provides early feedback on a research idea/method. This feedback is offered regardless of whether a submission is accepted, and was very valuable. The MSR survey confirmed this. It is a form of research mentoring or shepherding that combines the feedback of peer review before costly data collection. Nonetheless, some authors remain wedded to their approach (as is their prerogative), and do not change to match what reviewers asked for (these submissions are usually rejected at Stage 1).

\item [RRs prevent research problems] RR pre-register analysis approaches. Registration is largely independent of the journal; one could simply register an analysis on the Open Science Foundation or ArXiv, with no requirement to get IPA, or approval from a journal to register. This is what registration is used for in the conventional narrative. This \emph{preregistration} commits the researchers to a particular analysis path and data selection, ahead of seeing the actual data. Then, the final results are published in a journal in the conventional manner. 

\item[RRs act as first-round review incentive] RR serves as in principle acceptance for publishing in a prestigious journal. The RR process is focused only on ``accepted" registrations, and offering quicker publication in a journal as a carrot (partly to encourage avoiding research bias, the second point). It also ensures the focus is properly on the importance of the question and the suitability of the methods used to detect it, rather than the results themselves.
\end{description}

\subsection{To What Does Registration Apply?}
Managing a RR track or special issue means grappling with the broad scope of software research methods. In fields such as psychology, research methods seem more standardized and have been developed (and argued about) for decades. Software engineering research, by contrast, is more interdisciplinary, and relies on methods from engineering, business, psychology, sociology, mathematics, physics, to name but a few. 
These methods can have feature a variety of data types, from continuous floating point simulation results to free-form qualitative text. 
They can be part of confirmatory research or exploratory research. 
RRs tend to support the former approach more readily.
Finally, software engineering researchers come from a host of different philosophical perspectives, although post-positivist paradigms dominate. 

Two types of submission in particular challenged our reviewers. In a \textbf{qualitative} study it is not common to take a philosophical perspective that is exploratory and knowledge-seeking. Such a protocol might propose one particular study approach, but then change that study approach as interview participants (for example) contradict assumptions. 
Reviewers in SE are often less familiar with qualitative approaches, so analysing a qualitative approach such as grounded theory or systematic reviews can be superficial (``how will you assess coding reliability") than an equivalent quantitative study such as a controlled experiment. 
However, researchers still benefit from the early feedback on the approach, for example, on the coding approach or sampling strategies \citep{Karhulahti2022}.

\textbf{Data mining} studies were also hard to review as Stage 1 proposals.
This again seems to be related to the degree of exploration the research proposed.
Data mining studies either apply existing ML algorithms (naive bayes, support vector machines, etc.) or propose new algorithms to a feature-engineered dataset (such as the NASA datasets~\citep{seacraftrepo}). 
The goal of these studies is to derive new insights about suitable features, the best performing learner, and new approaches to algorithm efficiency or accuracy. 
An example is applying a machine learning approach to bug localization~\cite{Heiden2019AnEO}.
Data mining studies are a large part of the SE research landscape, but do not typically specify confirmatory hypotheses {\em a priori}. 
For example, it would be unusual to see a claim that a specific ML algorithm should work better on Mozilla bugs than Chrome bugs.
The epistemic objective is to work on novel features and algorithms to improved software engineering data analysis~\cite{Menzies2021}.

The common theme to both approaches is the distinction between exploratory research and confirmatory research. Certainly medical trials and controlled psychology experiments focus on confirming a well-formed hypothesis H, ideally ``severely testable", i.e., makes a very specific, tightly bound and testable claim: ``not only that H agrees with the data, but that with high probability, H would not have passed the test so well, were H false~\citep[p. 92]{Mayo2006}". 

In an exploratory approach, however, the study is asking questions and takes no position on what the results should be. Changing observations as results emerge is a key part of the inductive nature of the process. Registration still seems to work here, however: a Stage 1 submission garners useful feedback from experts (e.g., ``why not try to use this dataset as well"); the analysis approach can still be spelled out, which frankly is just good research design, independent of the publication of the results. However, it does suggest swift review of the IPA is more complex, because the analysis is highly dependent on the data. 

To reconcile this, MSR in 2021 and 2022 has been marking some submissions to the RR track as ``continuity acceptance", whereby the paper is accepted as a Stage 1 proposal, but not given in-principal acceptance, and requiring further in depth review. This idea has currently also been extended to ICSME and ESEM venues. 


\begin{table*}
    \begin{tabular}{p{5.5cm}p{5.5cm}}
    \toprule
         Benefits & Disadvantages \\ \midrule
         Shareable protocols for research replication. & More effort from researchers. \\
         Focus is on research, not publication. & Limited acceptance by journals so far. \\
         Improved rigour in reporting.  & Rigour can mean different things to different people/communities \citep{Storey2020TheWW}. \\
         Early peer review on research approach. & Not all research strategies are registerable.\\ \bottomrule
    \end{tabular}
    \caption{Benefits and Disadvantages of Registered Reports in SE.}
    \label{tab:benefits}
    \end{table*}

\subsection{Future Directions for RR and Open Questions}
Registered reports are very new in software engineering. Many questions remain. Foremost in our minds are the following:

\begin{itemize}
    \item What does pre-registration look like in qualitative research, or epistemologies which differ from post-positivism? What if statistical frameworks are not applicable? Such registrations might focus on early feedback as in a doctoral symposium or the Work In Progress sessions at ICER.\footnote{\url{https://computinged.wordpress.com/2019/05/31/come-hang-out-with-wil-and-me-to-talk-about-new-research-ideas-acm-icer-2019-work-in-progress-workshop/}}
    \item Does it make sense to support \emph{exploratory research} in a pre-registered setting? What are the advantages? The current thinking is that deviation from the initial protocol is tolerated if the deviation is small and not based on looking at the data (for example, changing a statistical test to non-parameteric). But more purely exploratory work may not be a good fit for registration~\citep{Waldron2022NotAllPre}, and should not be seen as ``less than'' because of this. 
    \item What is the quality of the final paper, and is in-principle acceptance at Stage 1 sufficiently rigorous? To date the EMSE papers have nearly always had major revisions to the Stage 2 submission, as reviewers emphasize rigour and the community adjusts to the model. This emphasis, however, might mean an RR process results in longer publication time (however, the starting point for reviews---research design---is earlier as well).
    \item How common are unreplicable results and researcher bias in software engineering anyway? Do we also have problems with suspiciously large numbers of studies with p-values close to 0.05? Studies to date have shown a lack of statistical maturity~\citep{Neto2019EvolutionOS} which precludes even answering such a question. Another study shows puzzling lack of retractiions in ACM and IEEE publications~\cite{}. 
    \item Can we better connect conference and journal review management systems to facilitate the administration of registered reports? The open scholarship community has numerous platforms for hosting preprints and protocols, such as PeerCommunityIn\footnote{\url{rr.peercommunityin.org}}, AsPredicted\footnote{\url{https://aspredicted.org}}, and the Open Science Foundation Platforms. 
\end{itemize}
The ultimate question is whether registered reports {\em help or hurt} the quality of research in software engineering. We hope to analyze this question as the community publishes more registered reports. In the meantime, we are strongly encouraged by the interest from the community and the many benefits of RR we have observed.

\section*{Acknowledgments}
Thanks to our collaborators in getting RR in software engineering started, including Prem Devanbu and the MSR steering committee, Janet Siegmund, Tim Menzies, Christoph Treude, Tegawendé Bissayandé, David Lo, Jeff Carver, as well as the many reviewers who have helped out. Thanks to David Lo and Martin Shepperd for helpful reviews on the manuscript. 

Without Tom Zimmermann and Robert Feldt, editors-in-chief at the Journal of Empirical Software Engineering, none of this would be possible. A final thanks to the folks at the Open Science Foundation and Peer Community In Registered Reports for pushing open science in general and making RR materials easily accessible.

\bibliography{rr}

\end{document}